\documentclass[12pt]{article}

\usepackage{amsthm}
\theoremstyle{plain}

\theoremstyle{definition}

\theoremstyle{proposition}

\theoremstyle{lemma}

\theoremstyle{remark}

\usepackage{amsmath}
\usepackage{amssymb}
\usepackage[dvips]{graphicx}
\usepackage{cite}

\makeatletter
 
  \@addtoreset{equation}{section}
 \makeatother

\begin{document}
\setlength{\oddsidemargin}{0cm}
\setlength{\baselineskip}{7mm}

\begin{titlepage}

~~\\

\vspace*{0cm}
    \begin{Large}
       \begin{center}
         {Quantum Theory of the Universe Based on Bayesian Probability}
       \end{center}
    \end{Large}
\vspace{1cm}

\begin{center}
           Matsuo S{\sc ato}\footnote
           {
e-mail address : msato@hirosaki-u.ac.jp}\\
      \vspace{1cm}
       
         {\it Department of Natural Science, Faculty of Education, Hirosaki University\\ 
 Bunkyo-cho 1, Hirosaki, Aomori 036-8560, Japan}

\end{center}

\hspace{5cm}

\begin{abstract}
\noindent
We  formulate a quantum theory of the Universe based on Bayesian probability. In this theory, the probability of the Universe is not a frequency probability, which can be obtained by observing experimental results several times,  but is a Bayesian probability, which can define a probability of an event that occurs just once. As an example, by applying the quantum theory of the Universe to an action of a scalar field theory in the four dimensions as a toy model for the theory of the Universe, we explicitly obtain the probability of the Universe and the action of matters in the Universe.
\end{abstract}

\vfill
\end{titlepage}
\vfil\eject

\setcounter{footnote}{0}

\section{Introduction}
\setcounter{equation}{0}
The quantum theory of matters is defined by a path integral  $\int \mathcal{D}\phi e^{-S[\phi]}$  with  an action $S[\phi]$, which represents the total probability of  a canonical distribution $e^{-S[\phi]}$. That is, the theory is defined by a standard framework of probability theory, so called frequency probability, which is defined based on populations, because experiments on matters can be made several times in  laboratory systems.  On the other hand,  the time evolution of the Universe cannot repeat  or cannot be made experiments on. Thus, we cannot define a quantum theory of the Universe by using frequency probability. Therefore, we propose to define a quantum theory of the Universe based on Bayesian probability,\footnote{Bayesian probability is a base of the machine learning recently developed extensively. See \cite{Bishop, Data, Watanabe} as examples of text books for Bayesian probability.} which can give a probability of an event that occurs just once.\footnote{For example, Bayesian probability gives a probability of a state of  genes in a family tree, from an information of diseases in it. A state of  genes in a family tree occurs just once. }

Bayes' theorem states that 
$p(\bar{\phi}|\tilde{\phi})=p(\tilde{\phi}|\bar{\phi})p(\bar{\phi})/Z$, 
where 
$Z=\int d\bar{\phi} p(\tilde{\phi}|\bar{\phi})p(\bar{\phi})$. $p(x)$ is a probability that an event $x$ occurs, and 
$p(x|y)$ is a conditional probability that an event $x$ occurs under the condition that an event $y$ occurs. Bayes' theorem can be proven easily in the framework of the frequency probability as follows. In this framework, all the probabilities are defined by populations and $p(x|y):=\frac{p(x, y)}{p(y)}=\frac{p(x, y)}{\int d x p(x,y)}$, where $p(x,y)$ is a probability that events $x$ and $y$ occur. Actually, by using this, we can prove it as $p(\bar{\phi}|\tilde{\phi})=\frac{p(\bar{\phi},\tilde{\phi})}{p(\tilde{\phi})}=\frac{p(\bar{\phi},\tilde{\phi})}{\int d\bar{\phi} p(\bar{\phi},\tilde{\phi})}=p(\tilde{\phi}|\bar{\phi})p(\bar{\phi})/Z$. On the other hand, in the framework of Bayesian probability, the definitions are different: so-called posterior distribution $p(\bar{\phi}|\tilde{\phi})$, which is an analogue of the  conditional probability $p(\bar{\phi}|\tilde{\phi})$ in the frequency probability, is defined by the right-hand side of $p(\bar{\phi}|\tilde{\phi}):=p(\tilde{\phi}|\bar{\phi})p(\bar{\phi})/Z$, where $p(\bar{\phi})$  and $p(\tilde{\phi}|\bar{\phi})$ are called prior distribution and statistical model,\footnote{$p(x|y)$ is automatically normalized as $\int dx p(x|y)=1$ by definition in the frequency probability. Thus, the statistical model must be normalized as well.} respectively, defined without a population. That is, the posterior distribution $p(\bar{\phi}|\tilde{\phi})$ is defined even if a population is not defined or we cannot observe the conditional probability that  the event $\bar{\phi}$ occurs under the condition that the event $\tilde{\phi}$ occurs.

A summary of the application of the Bayesian probability to the Universe is as follows. $\bar{\phi}$ is a background that represents an Universe, whereas $\tilde{\phi}$ is a fluctuation that represents matters.  For example, in a theory of gravity,  $\bar{\phi}$ represents a space-time itself and a vacuum expectation value of matters, whereas  $\tilde{\phi}$ represents a graviton and matters. An action of a theory of the Universe gives a statistical model  $p(\tilde{\phi}|\bar{\phi})$ representing a probability that  $\tilde{\phi}$ around $\bar{\phi}$ occurs.  The posterior probability represents a probability of the Universe when fluctuations occur. This is defined although it cannot be observed directly. In section 2, we apply the Bayesian probability to the Universe precisely and formulate a quantum theory of the Universe.  In section 3, we explicitly obtain the probability of the Universe and the action of matters in the Universe in the case that we treat an action of a scalar field theory in the four dimensions as a toy model of a theory of the Universe. 

\section{Formalism}
\setcounter{equation}{0}
A prior distribution is given in general as $p(\bar{\phi}):=e^{-S_p[\bar{\phi}]}$, where $\bar{\phi}$ is a background representing an Universe and $S_p[\bar{\phi}]$ is an arbitrary function of $\bar{\phi}$ that is normalized as $\int \mathcal{D}\bar{\phi} e^{-S_p[\bar{\phi}]}=1$. A probability of the Universe is not the prior distribution itself but is obtained from the posterior distribution. Moreover, we will see that the probability of the Universe does not depend on the prior distribution later. After all, we can set the prior distribution 1 ($S_p[\bar{\phi}]=0$). 

By the quantum theory of matters, a probability of a correlation of N points $ f_i(\tilde{\phi})$ ($i =1, \cdots, N$) in an experiment in a background $\bar{\phi}$ is given by
\begin{equation}
<\prod_{i=1}^{N} f_i(\tilde{\phi})>
=\int \mathcal{D} \tilde{\phi}
\frac{1}{Z[\bar{\phi}]}e^{-S[\bar{\phi}+\tilde{\phi}]}
\prod_{i=1}^{N} f_i(\tilde{\phi}),
\end{equation}
where $S[\bar{\phi}+\tilde{\phi}]$ is given by expanding the action of a theory of the Universe $S[\phi]$ with respect to $\tilde{\phi}$ around $\bar{\phi}$, and $Z[\bar{\phi}]:=\int \mathcal{D} \tilde{\phi} e^{-S[\bar{\phi}+\tilde{\phi}]}$. This means that a probability that a path $\tilde{\phi}$ in a path integral occurs, is given by $\frac{1}{Z[\bar{\phi}]}e^{-S[\bar{\phi}+\tilde{\phi}]}$. Thus, we define a statistical model where a single fluctuation around a fixed background occurs as 
$p(\tilde{\phi}|\bar{\phi}):=\frac{1}{Z[\bar{\phi}]}e^{-S[\bar{\phi}+\tilde{\phi}]}
=e^{-S_0[\bar{\phi}, \tilde{\phi}]}$, where 
$S_0[\bar{\phi}, \tilde{\phi}]:=S[\bar{\phi}+\tilde{\phi}]+\log (Z[\bar{\phi}])$.

Next, we generalize the statistical model to a model that has multiple fluctuations. In another experiment, a probability of a correlation of $N'$ points $g_j(\tilde{\phi}')$  ($j =1, \cdots, N'$) in the same background $\bar{\phi}$, is given by $<\prod_{j=1}^{N'} g_j(\tilde{\phi}')>'$. Then,  a probability that we have correlation of $N'$ points in an experiment and a correlation of $N'$ points in another experiment is given by the product 
\begin{equation}
<\prod_{i=1}^{N} f_i(\tilde{\phi})><\prod_{j=1}^{N'} g_j(\tilde{\phi}')>'
=\int \mathcal{D} \tilde{\phi}\int \mathcal{D} \tilde{\phi}' 
e^{-S_0[\bar{\phi}, \tilde{\phi}]}e^{-S_0[\bar{\phi}, \tilde{\phi}']}
\prod_{i=1}^{N} f_i(\tilde{\phi})\prod_{j=1}^{N'} g_j(\tilde{\phi}').
\end{equation}
This means that a probability that a path $\tilde{\phi}$ in an experiment and another path $\tilde{\phi}'$ in another experiment occurs, is given by $e^{-S_0[\bar{\phi}, \tilde{\phi}]}e^{-S_0[\bar{\phi}, \tilde{\phi}']}$.  Thus, we define a statistical model where two independent fluctuations around a fixed background occur as $p(\tilde{\phi}|\bar{\phi}):=e^{-S_0[\bar{\phi}, \tilde{\phi}]}e^{-S_0[\bar{\phi}, \tilde{\phi}']}$. In the same way,  we define a statistical model where $N$ independent fluctuations around a fixed background occur as  $p(\tilde{\phi}_1, \cdots, \tilde{\phi}_N|\bar{\phi})=\prod_{i=1}^Np(\tilde{\phi}_i|\bar{\phi})
=e^{-\sum_{i=1}^N S_0[\bar{\phi}, \tilde{\phi}_i]}.$

A posterior distribution that a Universe $\bar{\phi}$ occurs under the condition that independent matters $\tilde{\phi}_1, \cdots, \tilde{\phi}_N$ occur, is defined as 
\begin{eqnarray}
p(\bar{\phi}|\tilde{\phi}_1, \cdots, \tilde{\phi}_N)
=
\frac{1}{Z_0(\tilde{\phi}_1, \cdots, \tilde{\phi}_N)}p(\tilde{\phi}_1, \cdots, \tilde{\phi}_N|\bar{\phi})
p(\bar{\phi})
=
\frac{1}{Z_0(\tilde{\phi}_1, \cdots, \tilde{\phi}_N)}e^{-\sum_{i=1}^N S_0[\bar{\phi}, \tilde{\phi}_i]-S_p[\bar{\phi}]}, \nonumber \\
\end{eqnarray}
where
$Z_0(\tilde{\phi}_1, \cdots, \tilde{\phi}_N):=\int \mathcal{D}\bar{\phi}p(\tilde{\phi}_1, \cdots, \tilde{\phi}_N|\bar{\phi}) p(\bar{\phi})
=
\int \mathcal{D} \bar{\phi} e^{-\sum_{i=1}^N S_0[\bar{\phi}, \tilde{\phi}_i]-S_p[\bar{\phi}]}$.

We are going to obtain a probability of the Universe by taking large $N$ expansion of the posterior distribution. Matters $\tilde{\phi}_1, \cdots, \tilde{\phi}_N$ are subject to the same probability, called true probability $q(\tilde{\phi})=e^{-S_t [ \tilde{\phi}]}$, where matters occur in the Universe. That is, $S_t [\tilde{\phi}]$ is the action of matters in the Universe. The normalization is given by $\int \mathcal{D}\tilde{\phi} e^{-S_t [ \tilde{\phi}]}=1$. True probability will be determined at last. 

We define an effective action as $K_0(\bar{\phi}):= \int \mathcal{D} \tilde{\phi} q(\tilde{\phi}) S_0[\bar{\phi}, \tilde{\phi}]$. We name $\bar{\phi}$ that gives the minimum of $K_0(\bar{\phi})$, $\bar{\phi}_0$. Then, $\bar{\phi}_0$ satisfies 
\begin{equation}
\frac{d K_0(\bar{\phi})}{d \bar{\phi}_m}|_{\bar{\phi}=\bar{\phi}_0}=0 \quad (m=1, \cdots, d). \label{extremal}
\end{equation}
In order to take large $N$ expansion of the posterior distribution, we define a deformed action whose effective action has minimum value 0, as 
$S_n[\bar{\phi}, \tilde{\phi}]
:=
S_0[\bar{\phi}, \tilde{\phi}]-S_0[\bar{\phi}_0, \tilde{\phi}].$
Actually, $S_n[\bar{\phi}_0, \tilde{\phi}]=0$ and the effective action 
$K(\bar{\phi})
:= 
\int \mathcal{D} \tilde{\phi} q(\tilde{\phi}) S_n[\bar{\phi}, \tilde{\phi}]
=
K_0(\bar{\phi})- \int \mathcal{D} \tilde{\phi} q(\tilde{\phi}) S_0[\bar{\phi}_0, \tilde{\phi}]$ 
satisfies 
\begin{eqnarray}
K(\bar{\phi}_0)&=&0 \label{Taylor1} \\ 
\frac{d K(\bar{\phi})}{d \bar{\phi}_m}|_{\bar{\phi}=\bar{\phi}_0}&=&0, \label{Taylor2}
\end{eqnarray}
because 
$\frac{d K(\bar{\phi})}{d \bar{\phi}_m}=\frac{d K_0(\bar{\phi})}{d \bar{\phi}_m}$. In addition, the posterior distribution defined by the deformed action coincides with that defined by the original action:
$p(\bar{\phi}|\tilde{\phi}_1, \cdots, \tilde{\phi}_N)=
\frac{1}{Z(\tilde{\phi}_1, \cdots, \tilde{\phi}_N)}
e^{-\sum_{i=1}^N S_n[\bar{\phi}, \tilde{\phi}_i]-S_p[\bar{\phi}]},
$
where 
\begin{equation}
Z(\tilde{\phi}_1, \cdots, \tilde{\phi}_N):=
\int \mathcal{D} \bar{\phi} e^{-\sum_{i=1}^N S_n[\bar{\phi}, \tilde{\phi}_i]-S_p[\bar{\phi}]}=e^{\sum_{i=1}^N S_0[\bar{\phi}_0, \tilde{\phi}_i]}
Z_0(\tilde{\phi}_1, \cdots, \tilde{\phi}_N). \label{PartitionFunction}
\end{equation}

Because the expectation value of $\sum_{j=1}^N S_n[\bar{\phi}, \tilde{\phi}_j]$ in (\ref{PartitionFunction}) is given by 
$\prod_{i=1}^N \int \mathcal{D} \tilde{\phi}_i q(\tilde{\phi}_i) \sum_{j=1}^N S_n[\bar{\phi}, \tilde{\phi}_j]
=
N K(\bar{\phi})$,
we define its fluctuation as 
\begin{equation}
\eta(\bar{\phi}|\tilde{\phi}_1, \cdots, \tilde{\phi}_N)
:=
\frac{1}{\sqrt{N}}(N K(\bar{\phi}) - \sum_{i=1}^N S_n[\bar{\phi}, \tilde{\phi}_i]). \label{eta}
\end{equation} 
$\eta(\bar{\phi}|\tilde{\phi}_1, \cdots, \tilde{\phi}_N)$ converges to a Gaussian process in $N \to \infty$ by the central limit theorem with respect to the random valuables $S_n[\bar{\phi}, \tilde{\phi}_i]$. The expectation value and the correlation functions are given by
$E(\bar{\phi})
:=
\prod_{i=1}^N \int \mathcal{D} \tilde{\phi}_i q(\tilde{\phi}_i)
\eta(\bar{\phi}|\tilde{\phi}_1, \cdots, \tilde{\phi}_N)
=
0$ 
and
\begin{eqnarray}
\sigma(\bar{\phi}_1, \bar{\phi}_2)
&:=&
\prod_{i=1}^N \int \mathcal{D} \tilde{\phi}_i q(\tilde{\phi}_i)
\eta(\bar{\phi}_1|\tilde{\phi}_1, \cdots, \tilde{\phi}_N)
\eta(\bar{\phi}_2|\tilde{\phi}_1, \cdots, \tilde{\phi}_N) \nonumber  \\
&=&
\int \mathcal{D} \tilde{\phi} q(\tilde{\phi})
S_n[\bar{\phi}_1, \tilde{\phi}]
S_n[\bar{\phi}_2, \tilde{\phi}]
-
K(\bar{\phi}_1) 
K(\bar{\phi}_2).
\end{eqnarray}
We also have
\begin{equation}
\eta(\bar{\phi}_0|\tilde{\phi}_1, \cdots, \tilde{\phi}_N)=0.
\label{Taylor3}
\end{equation}
We abbreviate $\eta(\bar{\phi}|\tilde{\phi}_1, \cdots, \tilde{\phi}_N)$ to $\eta(\bar{\phi})$ in the following. 
The partition function can be written in terms of $K(\bar{\phi})$ and $\eta(\bar{\phi})$ by definition of $\eta(\bar{\phi})$:
$Z(\tilde{\phi}_1, \cdots, \tilde{\phi}_N):=
\int \mathcal{D} \bar{\phi} e^{-NK(\bar{\phi}) +\sqrt{N}\eta(\bar{\phi})-S_p[\bar{\phi}]}$.

We divide the partition function to the main part 
\begin{equation}
Z^{(1)}(\tilde{\phi}_1, \cdots, \tilde{\phi}_N):=\int_{K(\bar{\phi})<\epsilon_N} \mathcal{D} \bar{\phi} e^{-NK(\bar{\phi}) +\sqrt{N}\eta(\bar{\phi})-S_p[\bar{\phi}]}
\end{equation}
and the remaining part
\begin{equation}
Z^{(2)}(\tilde{\phi}_1, \cdots, \tilde{\phi}_N):=\int_{K(\bar{\phi})>\epsilon_N} \mathcal{D} \bar{\phi} e^{-NK(\bar{\phi}) +\sqrt{N}\eta(\bar{\phi})-S_p[\bar{\phi}]}:
\end{equation}
\begin{equation}
Z(\tilde{\phi}_1, \cdots, \tilde{\phi}_N)=Z^{(1)}(\tilde{\phi}_1, \cdots, \tilde{\phi}_N)+Z^{(2)}(\tilde{\phi}_1, \cdots, \tilde{\phi}_N),
\end{equation}
where $\epsilon_N$ satisfies
\begin{eqnarray}
\lim_{N \to \infty} \epsilon_N&=&0  \label{zero}\\
\lim_{N \to \infty} \sqrt{N} \epsilon_N&=& \infty. 
\label{infty}
\end{eqnarray}
$Z^{(2)}(\tilde{\phi}_1, \cdots, \tilde{\phi}_N)
=
O(e^{-\sqrt{N}})$ because
\begin{eqnarray}
&&Z^{(2)}(\tilde{\phi}_1, \cdots, \tilde{\phi}_N) \nonumber \\
&\le&
 e^{-N\epsilon_N +\sqrt{N} \sup_{\bar{\phi}} \eta(\bar{\phi})}
\int_{K(\bar{\phi})>\epsilon_N} \mathcal{D} \bar{\phi}
e^{-S_p[\bar{\phi}]} \nonumber \\
&\le&
 e^{-\sqrt{N} (\frac{\sqrt{N} \epsilon_N}{2} -\frac{( \sup_{\bar{\phi}} \eta(\bar{\phi}))^2}{2\sqrt{N} \epsilon_N})}
\int \mathcal{D} \bar{\phi}
e^{-S_p[\bar{\phi}]} \nonumber \\
&=&
O(e^{-\sqrt{N}}),
\end{eqnarray}
where we have used the inequality of additive geometric mean,
\begin{equation}
\sqrt{N} \sup_{\bar{\phi}} \eta(\bar{\phi}) 
\le 
\frac{1}{2}(N \epsilon_N + \frac{( \sup_{\bar{\phi}} \eta(\bar{\phi}))^2}{\epsilon_N}).
\end{equation}

We evaluate the main part. In the following, we treat only the case that $\bar{\phi}_0$ is unique and $J_{mn}(\bar{\phi}):=\frac{\partial}{\partial \bar{\phi}_m}\frac{\partial}{\partial \bar{\phi}_n} K(\bar{\phi})$ is regular, namely its eigenvalues are all positive. In the general case, we are working on evaluating the main part in progress \cite{progress}. 
From (\ref{Taylor1}), (\ref{Taylor2}), (\ref{Taylor3}),  and the mean value theorem, there exist $\bar{\phi}'$ and $\bar{\phi}''$ in $||\bar{\phi}'-\bar{\phi}_0|| \le ||\bar{\phi}-\bar{\phi}_0||$ and 
$||\bar{\phi}''-\bar{\phi}_0|| \le ||\bar{\phi}-\bar{\phi}_0||$ such that
\begin{eqnarray}
K(\bar{\phi})
&=&
\frac{1}{2}J_{mn}(\bar{\phi}')(\bar{\phi}_m-\bar{\phi}_{0 m})(\bar{\phi}_n-\bar{\phi}_{0 n})
\label{expand}
\\
\eta(\bar{\phi})
&=&\xi_m(\bar{\phi}'')
(\bar{\phi}_m-\bar{\phi}_{0 m}),
\end{eqnarray}
where
$\xi_m(\bar{\phi})
:=
\frac{\partial}{\partial \bar{\phi}_m}
\eta(\bar{\phi})$.
Because the integration region of the main part is\begin{equation}
K(\bar{\phi})<\epsilon_N
\label{epsilon}
\end{equation}
and (\ref{zero}),
we have
$\bar{\phi}'=\bar{\phi}_0+o(1)$
and
$\bar{\phi}''=\bar{\phi}_0+o(1)$.
Thus, the integrand of the main part is given by 
\begin{eqnarray}
&&e^{-NK(\bar{\phi}) +\sqrt{N}\eta(\bar{\phi})-S_p[\bar{\phi}]} \nonumber \\
&=&
e^{-\frac{N}{2} J_{mn}(\bar{\phi}_0)(\bar{\phi}_m-\bar{\phi}_{0 m})(\bar{\phi}_n-\bar{\phi}_{0 n})
+\sqrt{N} \xi_m(\bar{\phi}_0)
(\bar{\phi}_m-\bar{\phi}_{0 m})-S_p[\bar{\phi}_0]}(1+o(1)) \nonumber \\
&=&
e^{-\frac{N}{2}  (J^{\frac{1}{2}}_{mn}(\bar{\phi}_0)(\bar{\phi}_n-\bar{\phi}_{0 n}-\frac{1}{\sqrt{N}}J^{-1}_{nl}\xi_l(\bar{\phi}_0)))^2
+\frac{1}{2}(J^{-\frac{1}{2}}_{mn}(\bar{\phi}_0)\xi_n(\bar{\phi}_0))^2-S_p[\bar{\phi}_0]}(1+o(1)),
\end{eqnarray}
which is Gaussian with a dispersion
$\frac{1}{\sqrt{N}}J_{mn}^{-\frac{1}{2}}$, 
where
$J_{mn}^{\frac{1}{2}}J_{nl}^{\frac{1}{2}}=J_{ml}$.

From  (\ref{expand}) and (\ref{epsilon}) and the regularity of $J_{mn}(\bar{\phi}_0)$, the integration region is $\sqrt{\epsilon_N} \gtrsim |\bar{\phi}_m-\bar{\phi}_{0m}|$.
From (\ref{infty}), we have $\epsilon_N \gtrsim \frac{1}{\sqrt{N}}$, and then $\sqrt{\epsilon_N} \gtrsim \frac{1}{N^{\frac{1}{4}}} > \frac{1}{\sqrt{N}}$. Thus, the integration region $\sqrt{\epsilon_N}$ covers the center $\frac{1}{\sqrt{N}}J^{-1}_{ml}\xi_l(\bar{\phi}_0)$ plus and minus the deviation $\frac{1}{\sqrt{NJ_m}}$, where $J_m$ are eigenvalues of $J_{mn}$. Therefore,
\begin{eqnarray}
&&Z^{(1)}(\tilde{\phi}_1, \cdots, \tilde{\phi}_N) \nonumber \\
&=&\int \mathcal{D} \bar{\phi} 
e^{-\frac{N}{2}  (J^{\frac{1}{2}}_{mn}(\bar{\phi}_0)(\bar{\phi}_n-\bar{\phi}_{0 n}-\frac{1}{\sqrt{N}}J^{-1}_{nl}\xi_l(\bar{\phi}_0)))^2
+\frac{1}{2}(J^{-\frac{1}{2}}_{mn}(\bar{\phi}_0)\xi_n(\bar{\phi}_0))^2-S_p[\bar{\phi}_0]}(1+o(1)) \nonumber \\
&=&(\frac{2\pi}{N})^{\frac{d}{2}}(detJ)^{-\frac{1}{2}}
e^{\frac{1}{2}(J^{-\frac{1}{2}}_{mn}(\bar{\phi}_0)\xi_n(\bar{\phi}_0))^2-S_p[\bar{\phi}_0]}(1+o(1))
\nonumber \\
&=&O(N^{-\frac{d}{2}})
>>O(e^{-\sqrt{N}})= Z^{(2)}(\tilde{\phi}_1, \cdots, \tilde{\phi}_N),
\end{eqnarray}
and then, 
\begin{equation}
Z(\tilde{\phi}_1, \cdots, \tilde{\phi}_N) =Z^{(1)}(\tilde{\phi}_1, \cdots, \tilde{\phi}_N) +Z^{(2)}(\tilde{\phi}_1, \cdots, \tilde{\phi}_N)  \to Z^{(1)}(\tilde{\phi}_1, \cdots, \tilde{\phi}_N)  \quad (N \to \infty).
\end{equation}
Thus, large $N$ expansion of the posterior distribution, namely the probability of the Universe is given by
\begin{equation}
p(\bar{\phi}|\tilde{\phi}_1, \cdots, \tilde{\phi}_N)=
\frac{1}{\bar{Z}(\tilde{\phi}_1, \cdots, \tilde{\phi}_N)}
e^{-\frac{N}{2}  (J^{\frac{1}{2}}_{mn}(\bar{\phi}_0)(\bar{\phi}_n-\bar{\phi}_{0 n}-\frac{1}{\sqrt{N}}J^{-1}_{nl}\xi_l(\bar{\phi}_0)))^2}(1+o(1)), \label{Universe}
\end{equation}
where
\begin{equation}
\bar{Z}(\tilde{\phi}_1, \cdots, \tilde{\phi}_N) 
:=\int_{K(\bar{\phi})<\epsilon_N} \mathcal{D} \bar{\phi} 
e^{-\frac{N}{2}  (J^{\frac{1}{2}}_{mn}(\bar{\phi}_0)(\bar{\phi}_n-\bar{\phi}_{0 n}-\frac{1}{\sqrt{N}}J^{-1}_{nl}\xi_l(\bar{\phi}_0)))^2}.
\end{equation}

We obtain the expectation value of the statistical model of matters in a Universe $p(\tilde{\phi}|\bar{\phi})
=e^{-S_0[\bar{\phi}, \tilde{\phi}]}
=e^{-S_0[\bar{\phi}_0, \tilde{\phi}]
-\frac{\partial}{\partial \bar{\phi}_m}S_0[\bar{\phi}, \tilde{\phi}]|_{\bar{\phi}=\bar{\phi}_0}(\bar{\phi}_m-\bar{\phi}_{0 m})}
(1+o(1)), 
$ with respect to the probability of the Universe (\ref{Universe}),
\begin{eqnarray}
p_N(\tilde{\phi}; \xi)
&=&
\int_{K(\bar{\phi})<\epsilon_N} \mathcal{D} \bar{\phi} p(\tilde{\phi}|\bar{\phi})p(\bar{\phi}|\tilde{\phi}_1, \cdots, \tilde{\phi}_N) \nonumber \\
&=&
 e^{-S_0[\bar{\phi}_0, \tilde{\phi}]
-\frac{1}{\sqrt{N}}J^{-1}_{ml}\xi_l(\bar{\phi}_0)\frac{\partial}{\partial \bar{\phi}_m}S_0[\bar{\phi}, \tilde{\phi}]|_{\bar{\phi}=\bar{\phi}_0}
+\frac{1}{2N}(J^{-\frac{1}{2}}_{mn}(\bar{\phi}_0)\frac{\partial}{\partial \bar{\phi}_n}S_0[\bar{\phi}, \tilde{\phi}]|_{\bar{\phi}=\bar{\phi}_0})^2}(1+o(1))  \nonumber \\
&=&
 e^{-S_0[\bar{\phi}_0, \tilde{\phi}]
-\frac{1}{\sqrt{N}}J^{-1}_{m}\xi_m(\bar{\phi}_0)\frac{\partial}{\partial \bar{\phi}_m}S_0[\bar{\phi}, \tilde{\phi}]|_{\bar{\phi}=\bar{\phi}_0}
+\frac{1}{2N} J^{-1}_{m}(\bar{\phi}_0)(\frac{\partial}{\partial \bar{\phi}_m}S_0[\bar{\phi}, \tilde{\phi}]|_{\bar{\phi}=\bar{\phi}_0})^2}(1+o(1)),  \label{pN}\nonumber \\
\end{eqnarray}
where we have diagonalized $J_{mn}$ by changing the basis of $\phi_m$.

In order to obtain a finite and non-trivial $N \to \infty$ limit of the large $N$ expansion of $p_N(\tilde{\phi}; \xi)$ (\ref{pN}),  we take a double scaling limit $J_m \to 0$ with $J^2_m N$ fixed.   $J_m \to 0$ is realized by making parameters $\lambda_n$ ($n=1, \cdots, d$) approach to their critical values $\lambda^*_n$ simultaneously with $N \to \infty$. $\lambda^*$ is determined by
\begin{equation}
J_{m}(\lambda^*)=0. \label{J=0}
\end{equation}
By Taylor expanding $K(\bar{\phi})$ around $\bar{\phi}=\bar{\phi}_0$, the integrand of the partition function becomes
\begin{eqnarray}
&&e^{-NK(\bar{\phi}) +\sqrt{N}\eta(\bar{\phi})-S_p[\bar{\phi}]} \nonumber \\
&=& e^{-N\sum_{n=0}^{\infty}J^{(n)}_{m_1, \cdots, m_n}(\bar{\phi}_0)(\bar{\phi}_{m_1}-\bar{\phi}_{0 m_1} )\cdots (\bar{\phi}_{m_n}-\bar{\phi}_{0 m_n} )+\sqrt{N}\eta(\bar{\phi})-S_p[\bar{\phi}]},\end{eqnarray}
where
$J^{(n)}_{m_1, \cdots, m_n}(\bar{\phi})
:=
\frac{1}{n!}\frac{\partial}{\partial \bar{\phi}_{m_1}}\cdots\frac{\partial}{\partial \bar{\phi}_{m_n}}K(\bar{\phi})$, especially, $J^{(0)}(\bar{\phi}_0)=0$, $J^{(1)}_m(\bar{\phi}_0)=0$ and $J^{(2)}_{m,n}(\bar{\phi}_0)=2J_{mn}$.  From this formula,  one can see that the partition function and posterior distribution depend on  $J_{mn}$ only through the product $NJ_{mn}$, perturbatively. Thus, the large $N$ corrections from the main part of the partition function to $p_N(\tilde{\phi}; \xi)$ (\ref{pN}) come from expansions with respect to $\frac{1}{NJ_{mn}}$ and $\frac{1}{\sqrt{N}}$ only. By fixing $J^2_m N$ in the double scaling limit, these corrections vanish, and the corrections from the remaining part of the partition function also vanish because the corrections' dependence on $J_m$ is at most 
$O(e^{-\sqrt{NJ_m}})=O(e^{-N^{\frac{1}{4}}})$. 
We make $\lambda_n$ approach to $\lambda^*_n$ canonically as 
\begin{equation}
\lambda_n=e^{-\frac{t_n}{N^d}}\lambda^*_n, \label{approach}
\end{equation} 
where $t_n$ and $d$ are positive parameters. Because 
$J_{m}(\lambda)=-\frac{t_n}{N^d}\lambda^*_n\frac{\partial J_m}{\partial \lambda_n}|_{\lambda=\lambda^*}+o(\frac{1}{N^d})$, $J_m^2N$ is fixed if and only if $d=\frac{1}{2}$.  Therefore, we obtain a finite and non-trivial limit, 
\begin{eqnarray}
\lim p_N(\tilde{\phi}; \xi)
=
 e^{-S_0^*[\bar{\phi}_0, \tilde{\phi}]
+(t_n \lambda^*_n\frac{\partial J_m(\bar{\phi}_0)}{\partial \lambda_n}|_{\lambda=\lambda^*})^{-1}\xi_m^*(\bar{\phi}_0)\frac{\partial}{\partial \bar{\phi}_m}S_0^*[\bar{\phi}, \tilde{\phi}]|_{\bar{\phi}=\bar{\phi}_0}
},
\end{eqnarray}
where $*$ represents that $\lambda=\lambda^*$ is substituted.

The random valuables $S_n[\bar{\phi}, \tilde{\phi}_i]$ are not singular in $\lambda \to \lambda^*$, because $S_n[\bar{\phi}, \tilde{\phi}_i]$ do not depend on $J_{mn}$.  Thus,  the central limit theorem states that $\eta(\bar{\phi}|\tilde{\phi}_1, \cdots, \tilde{\phi}_N)$ converges to a Gaussian process in the double scaling limit by definition of (\ref{eta}). In the same way, its differential $\xi_m(\bar{\phi}|\tilde{\phi}_1, \cdots, \tilde{\phi}_N)=\frac{\partial}{\partial \bar{\phi}_m}\eta(\bar{\phi}|\tilde{\phi}_1, \cdots, \tilde{\phi}_N)$ also converges to a Gaussian process in this limit.  Its expectation value is given by 
$E^*(\bar{\phi})
:=
\lim \prod_{i=1}^N \int \mathcal{D} \tilde{\phi}_i q(\tilde{\phi}_i)
\frac{\partial}{\partial \bar{\phi}_m}
\eta(\bar{\phi}|\tilde{\phi}_1, \cdots, \tilde{\phi}_N)
=
0,$
whereas its correlation function is given by 
\begin{eqnarray}
\sigma_{mn}^*(\bar{\phi}_1, \bar{\phi}_2)
&:=&
\lim \prod_{i=1}^N \int \mathcal{D} \tilde{\phi}_i q(\tilde{\phi}_i)
\frac{\partial}{\partial \bar{\phi}_{1 m}}
\eta(\bar{\phi}_1|\tilde{\phi}_1, \cdots, \tilde{\phi}_N)
\frac{\partial}{\partial \bar{\phi}_{2 n}}
\eta(\bar{\phi}_2|\tilde{\phi}_1, \cdots, \tilde{\phi}_N) \nonumber  \\
&=&
\lim \int \mathcal{D} \tilde{\phi} q(\tilde{\phi})
\frac{\partial}{\partial \bar{\phi}_{1 m}}S_n[\bar{\phi}_1, \tilde{\phi}]
\frac{\partial}{\partial \bar{\phi}_{2 n}}S_n[\bar{\phi}_2, \tilde{\phi}]
-
\lim \frac{\partial}{\partial \bar{\phi}_{1 m}}K(\bar{\phi}_1) 
\frac{\partial}{\partial \bar{\phi}_{2 n}}K(\bar{\phi}_2).  \nonumber \\
\end{eqnarray}
Its dispersion is given by
$\sigma_{mn}^*(\bar{\phi}):=\sigma_{mn}^*(\bar{\phi}, \bar{\phi})$ and especially, at $\bar{\phi}=\bar{\phi}_0$,
\begin{eqnarray}
\sigma_{mn}^*(\bar{\phi}_0)&=&\lim \int \mathcal{D} \tilde{\phi} q(\tilde{\phi})
\frac{\partial}{\partial \bar{\phi}_{m}}S_n[\bar{\phi}, \tilde{\phi}]|_{\bar{\phi}=\bar{\phi}_0}
\frac{\partial}{\partial \bar{\phi}_{ n}}S_n[\bar{\phi}, \tilde{\phi}]|_{\bar{\phi}=\bar{\phi}_0} \nonumber \\
&=&\lim
\int \mathcal{D} \tilde{\phi} q(\tilde{\phi})
\frac{\partial}{\partial \bar{\phi}_{m}}S_0[\bar{\phi}+\tilde{\phi}]|_{\bar{\phi}=\bar{\phi}_0}
\frac{\partial}{\partial \bar{\phi}_{ n}}S_0[\bar{\phi}+\tilde{\phi}]|_{\bar{\phi}=\bar{\phi}_0}  \nonumber \\
&&-
\lim \frac{\partial}{\partial \bar{\phi}_{m}}\log Z[\bar{\phi}]|_{\bar{\phi}=\bar{\phi}_0}
\frac{\partial}{\partial \bar{\phi}_{ n}}\log Z[\bar{\phi}]|_{\bar{\phi}=\bar{\phi}_0}. 
\end{eqnarray}
Thus, $\xi^*_m(\bar{\phi}_0)$ is subject to the normal distribution $\mathcal{N}(0, \sigma_{mn}^*(\bar{\phi}_0))$, that is, the probability is given by 
$p_{\xi^*}(\xi^*)=\frac{1}{(2\pi)^{\frac{d}{2}}det^{\frac{1}{2}}(\sigma^*(\bar{\phi}_0))}\exp(-\frac{1}{2}\sigma^{* -1}_{mn}(\bar{\phi}_0)\xi^*_m \xi^*_n)$. The expectation value with respect to this probability is given by
\begin{eqnarray}
p^*(\tilde{\phi})
&:=&\int d\xi^* p_{\xi^*}(\xi^*) \lim p_N(\tilde{\phi}; \xi) \nonumber \\
&=&
 e^{-S_0^*[\bar{\phi}_0, \tilde{\phi}]
+\frac{1}{2}\sigma^*_{ml}(\bar{\phi}_0)(t_n \lambda^*_n\frac{\partial J_m(\bar{\phi}_0)}{\partial \lambda_n}|_{\lambda=\lambda^*})^{-1}\frac{\partial}{\partial \bar{\phi}_m}S_0^*[\bar{\phi}, \tilde{\phi}]|_{\bar{\phi}=\bar{\phi}_0}
(t_{n'} \lambda^*_{n'}\frac{\partial J_l(\bar{\phi}_0)}{\partial \lambda_{n'}}|_{\lambda=\lambda^*})^{-1}\frac{\partial}{\partial \bar{\phi}_l}S_0^*[\bar{\phi}, \tilde{\phi}]|_{\bar{\phi}=\bar{\phi}_0}}.   \label{p*} \nonumber \\
\end{eqnarray}
This is identified with the true probability\footnote{$p^*(\tilde{\phi})$ is automatically normalized: $\int \mathcal{D} \tilde{\phi}p^*(\tilde{\phi})=\int \mathcal{D} \tilde{\phi}\int d\xi^* p_{\xi^*}(\xi^*) \lim p_N(\tilde{\phi}; \xi)=\int d\xi^* p_{\xi^*}(\xi^*) \lim \int \mathcal{D} \bar{\phi} \int \mathcal{D} \tilde{\phi} p(\tilde{\phi}|\bar{\phi})p(\bar{\phi}|\tilde{\phi}_1, \cdots, \tilde{\phi}_N) =1$.} $q(\tilde{\phi})$: $q(\tilde{\phi})=p^*(\tilde{\phi})$\footnote{In practical Bayesian statistics,  the right-hand side is nearer left-hand side in a better statistical model. In a quantum theory of the Universe, the right-hand side should equal the left-hand side and the equation should be treated to determine the true probability, because the statistical model is obtained from the action of a theory of the Universe, that is, the model must be the best.}.
 $q(\tilde{\phi})$ is determined by solving this equation because (\ref{p*}) is a functional of $q(\tilde{\phi})$.

\section{Example}
\setcounter{equation}{0}
In this section, by treating the action of an Euclidean scalar field theory in the four dimensions $S[\phi]= \int d^4x (\frac{1}{2}(\partial_m \phi)^2 +V(\phi))$ as a toy model of the action of the theory of the Universe, we explicitly obtain the probability of the Universe and the true probability. In the following, we assume that $\bar{\phi}$ and $\bar{\phi}_0$ are constant and the fluctuations are up to the fourth order.

The action is expanded as $S[\bar{\phi}+\tilde{\phi}]= \int d^4x (\frac{1}{2}(\partial_m \tilde{\phi})^2 +V(\bar{\phi})+V'(\bar{\phi})\tilde{\phi}+\frac{1}{2}V''(\bar{\phi})\tilde{\phi}^2+\frac{1}{6}V'''(\bar{\phi})\tilde{\phi}^3+\frac{1}{24}V''''(\bar{\phi})\tilde{\phi}^4)$. The true probability is generally written as $q(\tilde{\phi})=e^{-S_t [ \tilde{\phi}]}$, where $S_t[\tilde{\phi}]= \int d^4x (\frac{1}{2}(\partial_m \tilde{\phi})^2 +a_0+a_1\tilde{\phi}+\frac{1}{2}a_2\tilde{\phi}^2+\frac{1}{6}a_3\tilde{\phi}^3+\frac{1}{24}a_4\tilde{\phi}^4)$. The normalization determines $a_0=\log\int \mathcal{D} \tilde{\phi} \exp(-\int d^4x (\frac{1}{2}(\partial_m \tilde{\phi})^2 +a_1\tilde{\phi}+\frac{1}{2}a_2\tilde{\phi}^2+\frac{1}{6}a_3\tilde{\phi}^3+\frac{1}{24}a_4\tilde{\phi}^4))$. In the following, we evaluate  path-integrals up to one-loop by using $\overline{MS}$ scheme\footnote{In the string geometry theory \cite{StringGeometry, TopologicalStringGeometry}, a perturbative string theory is obtained by expanding fluctuations up to the second order. In this case, we only consider the fluctuations up to the second order, and then the path-integral is exact.}.  We have 
$\log Z(\bar{\phi})=V_4(-V(\bar{\phi})+\frac{1}{2}\frac{V'(\bar{\phi})^2}{V''(\bar{\phi})}-\frac{1}{(8\pi)^2}V''(\bar{\phi})^2(\log\frac{V''(\bar{\phi})}{ M^2}-\frac{3}{2}))$.  The effective action is given by 
\begin{eqnarray}
K_0(\bar{\phi})&=&V_4(-\frac{a_1}{a_2}V'(\bar{\phi})+\frac{1}{2}((\frac{a_1}{a_2})^2+\frac{1}{16\pi^2}a_2(\log\frac{a_2}{M^2})-1))V''(\bar{\phi})-\frac{1}{32\pi^2}a_2^2(\log\frac{a_2}{M^2}-1)\nonumber \\
&&+\frac{1}{2}\frac{V'^2(\bar{\phi})}{V''(\bar{\phi})}-\frac{1}{64\pi^2}V''^2(\bar{\phi})(\log\frac{V''(\bar{\phi})}{ M^2}-\frac{3}{2})).
\end{eqnarray}
(\ref{extremal}) is equivalent to 
(i) $V'''(\bar{\phi}_0)=0$ and $V'(\bar{\phi}_0)=\frac{a_1}{a_2}V''(\bar{\phi}_0)$
or
\begin{eqnarray}
(ii) &&V''(\bar{\phi}_0)\log\frac{V''(\bar{\phi}_0)}{ M^2} \nonumber \\
&=&
V''(\bar{\phi}_0)
+32\pi^2(\frac{V'(\bar{\phi}_0)}{V'''(\bar{\phi}_0)}-\frac{a_1}{a_2}\frac{V''(\bar{\phi}_0)}{V'''(\bar{\phi}_0)}
-\frac{1}{2}\frac{V'^2(\bar{\phi}_0)}{V''^2(\bar{\phi}_0)}
+\frac{1}{2}(\frac{a_1}{a_2})^2+\frac{1}{32\pi^2}a_2(\log\frac{a_2}{M^2}-1)
). \nonumber \\
\end{eqnarray}
(i) or (ii) determines $\bar{\phi}_0$. $J(\bar{\phi}_0)$ and $\sigma(\bar{\phi}_0)$ are given by 
\begin{eqnarray}
J(\bar{\phi}_0)&=&V''(\bar{\phi}_0)-\frac{a_1}{a_2}V'''(\bar{\phi}_0)+V''''(\bar{\phi}_0)(\frac{1}{2}(\frac{a_1}{a_2})^2+\frac{1}{32\pi^2}a_2(\log\frac{a_2}{M^2}-1))-\frac{1}{2}\frac{V'^2(\bar{\phi}_0)V''''(\bar{\phi}_0)}{V''^2(\bar{\phi}_0)} \nonumber \\ 
&&-\frac{V'(\bar{\phi}_0)V'''(\bar{\phi}_0)}{V''(\bar{\phi}_0)}+\frac{V'^2(\bar{\phi}_0)V'''^2(\bar{\phi}_0)}{V''^3(\bar{\phi}_0)} +\frac{1}{32\pi^2}V''(\bar{\phi}_0)V''''(\bar{\phi}_0) \nonumber \\ 
&&-\frac{1}{32\pi^2}(V''''(\bar{\phi}_0)V''(\bar{\phi}_0)+V'''(\bar{\phi}_0)^2)\log\frac{V''(\bar{\phi}_0)}{ M^2}\label{Jexample} \\ 
\sigma(\bar{\phi}_0)&=&\frac{1}{2a_2}V''^2(\bar{\phi}_0) +\frac{a_1^2}{2a_2^3}V'''^2(\bar{\phi}_0) -\frac{2a_1}{a_2^2}V''(\bar{\phi}_0)V'''(\bar{\phi}_0) -\frac{1}{32\pi^2}V'''^2(\bar{\phi}_0)\log\frac{a_2}{M^2}.
\end{eqnarray}
The equation $q(\tilde{\phi})=p^*(\tilde{\phi})$, which determines $q(\tilde{\phi})$, is given up to the fourth order by
\begin{eqnarray}
a_1&=&V'_*(\bar{\phi}_0)-\sigma_*(\bar{\phi}_0)  (t \lambda^* \frac{dJ(\bar{\phi}_0)}{d\lambda}|_{\lambda=\lambda^*})^{-2}V''_*(\bar{\phi}_0)(V'_*(\bar{\phi}_0)-\frac{1}{2}\frac{V'^2_*(\bar{\phi}_0)V'''_*(\bar{\phi}_0)}{V''^2_*(\bar{\phi}_0)}-\frac{1}{32\pi^2}V''_*V'''_*(\log\frac{V''_*}{M^2}-1)) \nonumber \\
a_2&=&V''_*(\bar{\phi}_0)-\sigma_*(\bar{\phi}_0)  (t \lambda^* \frac{dJ(\bar{\phi}_0)}{d\lambda}|_{\lambda=\lambda^*})^{-2}(V''^2_*(\bar{\phi}_0)+V'''_*(\bar{\phi}_0)(V'_*(\bar{\phi}_0)-\frac{1}{2}\frac{V'^2_*(\bar{\phi}_0)V'''_*(\bar{\phi}_0)}{V''^2_*(\bar{\phi}_0)} \nonumber \\
&&-\frac{1}{32\pi^2}V''_*(\bar{\phi}_0)V'''_*(\bar{\phi}_0)(\log\frac{V''_*(\bar{\phi}_0)}{M^2}-1)) \nonumber \\
a_3&=&V_*'''(\bar{\phi}_0)-\sigma_*(\bar{\phi}_0)  (t \lambda^* \frac{dJ(\bar{\phi}_0)}{d\lambda}|_{\lambda=\lambda^*})^{-2}\nonumber \\
&&(3V_*''(\bar{\phi}_0)V_*'''(\bar{\phi}_0)+V_*''''(\bar{\phi}_0)
(V_*'(\bar{\phi}_0)-\frac{1}{2}\frac{V_*'^2(\bar{\phi}_0)V_*'''(\bar{\phi}_0)}{V_*''^2(\bar{\phi}_0)}-\frac{1}{32\pi^2}V_*''(\bar{\phi}_0)V_*'''(\bar{\phi}_0)(\log\frac{V_*''(\bar{\phi}_0)}{M^2}-1))) \nonumber \\
a_4&=&V_*''''(\bar{\phi}_0)-\sigma_*(\bar{\phi}_0)  (t \lambda^* \frac{dJ(\bar{\phi}_0)}{d\lambda}|_{\lambda=\lambda^*})^{-2}
(4V_*''(\bar{\phi}_0)V_*''''(\bar{\phi}_0)+3V_*'''^2(\bar{\phi}_0)), \label{LastEquation}
\end{eqnarray}
where $*$ represents that $\lambda^*$ determined by (\ref{J=0}) with (\ref{Jexample}) is substituted to $\lambda$ in $\bar{\phi}_0$, $\sigma$, $V$ and its derivatives. These equations determine $a_1$, $a_2$, $a_3$ and $a_4$.

We solve explicitly in case of  $V(\phi)=\frac{1}{4!}\lambda\phi^4+\frac{1}{2}m^2 \phi^2 + h \phi$. We only solve in case of (i) because there is no analytic solution in case of (ii). The solution to (i) is $\bar{\phi}_0=0$ and $\frac{a_1}{a_2}=\frac{h}{m^2}$. Then, we obtain $V'(\bar{\phi}_0)=h$, $V''(\bar{\phi}_0)=m^2$, $V'(\bar{\phi}_0)=0$ and $V'(\bar{\phi}_0)=\lambda$. The solution to  (\ref{J=0}) with (\ref{Jexample}) is $\lambda^*=\frac{32 \pi^2 m^2}{m^2(\log\frac{m^2}{M^2}-1)-a_2(\log\frac{a_2}{M^2}-1)}$. The solution to (\ref{LastEquation}) is $a_1=(\frac{1}{2} \pm \frac{\sqrt{t^2-2}}{2t})h$, $a_2=(\frac{1}{2} \pm \frac{\sqrt{t^2-2}}{2t})m^2$, $a_3=-\frac{1}{t^2 \pm t\sqrt{t^2-2}}\frac{h \lambda^*}{m^2}$ and $a_4=(1-\frac{4}{t^2 \pm t\sqrt{t^2-2}})\lambda^*$. When $t$ is the critical value $t=\sqrt{2}$, $a_1=\frac{1}{2}h$, $a_2=\frac{1}{2}m^2$, $a_3=-\frac{h \lambda^*}{2m^2}$, $a_4=-\lambda^*$ and $\lambda^*=\frac{64\pi^2}{\log\frac{m^2}{M^2}-\log(\frac{e}{2})}$. $a_4$ is positive because $\lambda^*$ is negative. We have obtained a stable theory of matters with $a_4=-\lambda^*>0$ from the theory of the Universe with $\lambda<0$ because $\lambda$ approaches to $\lambda^*$ as in  (\ref{approach}). One can confirm that $K_0(\bar{\phi})$ has the minimum at $\bar{\phi}=0$ by looking up globally. 

As a result, the probability of the Universe is (\ref{Universe}) with $J=m^2-\frac{\lambda}{64\pi^2}m^2(\log\frac{m^2}{M^2}-\log(\frac{e}{2}))$. The true probability, namely  the probability of matters in the Universe, is $q(\tilde{\phi})=\frac{1}{Z_t}\exp(-\int d^4x (\frac{1}{2}(\partial_m \tilde{\phi})^2 +\frac{1}{2}h\tilde{\phi}+\frac{1}{4}m^2\tilde{\phi}^2-\frac{h \lambda^*}{12m^2}\tilde{\phi}^3-\frac{1}{24}\lambda^*\tilde{\phi}^4))$, where $Z_t=\int \mathcal{D} \tilde{\phi} \exp(-\int d^4x (\frac{1}{2}(\partial_m \tilde{\phi})^2+\frac{1}{2}h\tilde{\phi}+\frac{1}{4}m^2\tilde{\phi}^2-\frac{h \lambda^*}{12m^2}\tilde{\phi}^3-\frac{1}{24}\lambda^*\tilde{\phi}^4))$. That is, the action of matters in the Universe is determined to be $\int d^4x (\frac{1}{2}(\partial_m \tilde{\phi})^2+\frac{1}{2}h\tilde{\phi}+\frac{1}{4}m^2\tilde{\phi}^2-\frac{h \lambda^*}{12m^2}\tilde{\phi}^3-\frac{1}{24}\lambda^*\tilde{\phi}^4))$, where $\lambda^*=\frac{64\pi^2}{\log\frac{m^2}{M^2}-\log(\frac{e}{2})}$.

\vspace{1cm}

\section{Conclusion and Discussion}
\setcounter{equation}{0}
In this paper, we  formulated a quantum theory of the Universe based on Bayesian probability. In this theory, the probability of the Universe is not a frequency probability, which can be obtained by observing experimental results several times,  but is a Bayesian probability, which can define a probability of an event that occurs just once. As an example, by applying the quantum theory of the Universe to an action of a scalar field theory in the four dimensions as a toy model for the theory of the Universe, we explicitly obtained the probability of the Universe and the action of matters in the Universe.

As an application, we will study to what extent the Universe is reproduced by applying the quantum theory of the Universe to the action of the Einstein gravity coupled with matters in the standard model, although it is not UV complete. Furthermore, it is necessary to apply the quantum theory of the Universe to the actions of candidates for the UV complete theory, such as matrix models, string field theories, loop quantum gravities, and string geometry theories\footnote{Especially, string geometry theory around a string geometry background reproduces a perturbative string theory on a string background \cite{StringGeometry, TopologicalStringGeometry}. Thus, it is important to apply the quantum theory of the Universe to the action of the string geometry theory and examine whether the action of the perturbative string theory in the Universe is uniquely determined and the Universe is reproduced including its origin.  Moreover, it is expected that string geometry theory is finite as a theory of the Universe because we only need to perform a path-integral over a very limited region as in (\ref{epsilon}).} to investigate the origin of the Universe.  



\vspace*{0cm}




\begin{thebibliography}{99}

\bibitem{Bishop}C.M.Bishop, ``Pattern Recognition and Machine Learning," Springer (2010)

\bibitem{Data}A.Gelman, J.B.Carlin, H.S.Stern, D.B.Dunson,A.Vehtari,D.B.Rubin, ``Bayesian Data Analysis,"  Chapman and Hall/CRC (2013)

\bibitem{Watanabe}S.Watanabe, ``Algebraic Geometry and Statistical Learning Theory," Cambridge University Press(2009)

\bibitem{progress}M.Sato, work in progress.


\bibitem{StringGeometry}M.Sato, ``String Geometry and Non-perturbative Formulation of String Theory," arXiv:1709.03506 [hep-th]

\bibitem{TopologicalStringGeometry}M.Sato, Y.Sugimoto, ``Topological String Geometry," arXiv:1903.05775 [hep-th]










\end{thebibliography}
\end{document}